# Multifunctional acoustic holography based on compact acoustic geometric-phase meta-array

Bingyi Liu[1,2], Qunshuo Wei[1], Zhaoxian Su[1], Yongtian Wang[1], and Lingling Huang[1*]

[1]*School of Optics and Photonics, Beijing Engineering Research Center of Mixed Reality and Advanced Display, Beijing Institute of Technology, Beijing, 100081, China*
[2]*Department of Physics, Paderborn University, Warburger Str. 100, 33098, Paderborn, Germany*

**Abstract**

Optical geometric-phase metasurface provides a robust and efficient means for light control by simply manipulating the spatial orientations of the in-plane anisotropic meta-atoms, where polarization conversion plays a vital role. However, the concept of acoustic geometric-phase modulation for acoustic field control remains unexplored because airborne acoustic waves lack a similar optical polarization conversion process. In this work, a new type of acoustic meta-atom with deep subwavelength feature size is theoretically investigated and further applied to acoustic field engineering based on the so-called acoustic geometric phase. Herein, tunable acoustic geometric-phase modulation of designated order is obtained via the near-field coupled orbital angular momentum transfer process, and the topological charge-multiplexed acoustic geometric phase endows our meta-arrays with multiple functionalities. Our work extends the capacity of acoustic meta-arrays in high-quality acoustic field reconstruction and offers new possibilities in multifunctional acoustic meta-holograms.

## I. INTRODUCTION

Metasurfaces, generally known as spatially patterned surfaces, have offered plethora applications in manipulating complex waves by using judiciously designed artificial scatters, i.e., so-called meta-atoms.[1,2] Notably, tailoring the incident wave through a designated local phase shift has been a cornerstone for numerous diffractive optical applications, such as meta-holograms[3-6] and meta-lenses.[7-10] Several phase shifting methods are well developed and widely implemented in optics, including the resonance phase, propagating phase, and, most intriguingly, geometric phase.[11] Geometric-phase metasurfaces (GPMs) gain their meticulous light manipulation capacity via the nontrivial geometric phase term acquired through the polarization evolutions, and continuous phase modulation covering the $2\pi$ range is available in both linear and nonlinear optics by simply rotating an in-plane anisotropic meta-atom. In short, the realization of GPMs is much simpler and more efficient than tuning the resonance or propagating phase, which relies on tailoring the dispersion properties of delicately designed meta-atoms. Therefore, it is of great research interest to extend the concept of optical GPMs to other disciplines, especially acoustics, seeking new possibilities in acoustic pressure field control.

Recently, the theory of acoustic spin[12-15] and acoustic spin-orbital angular momentum interaction[16] have been reported, which intrigues great research interest in exploring the acoustic counterparts of these fascinating and well-known concepts in optics. However, the analogous concept of the acoustic geometric phase has rarely been reported due to the absence of a similar polarization conversion process in acoustics, which should be crucial for its potential in obtaining a controllable phase shift by locally varying the structure orientation. Notably, the acoustic vortex of a given topological charge (TC) has been widely investigated for its powerful information processing capacity due to the orbital angular momentum (OAM) degree of freedom.[17,18] In particular, acoustic vortices propagating in either waveguides or free space have been well investigated with spiral phase plates made of acoustic metasurfaces, namely, acoustic vortex metasurfaces.[19-23] Interestingly, an acoustic vortex propagating in a helical acoustic waveguide has been revealed to support TC-dependent transport of acoustic waves, which, for the first time, verifies the existence of a spin-redirection geometric phase in acoustics.[24] This phenomenon can be regarded as an acoustic version of the geometric phase acquired by the structured light beam propagating through a helical optical path.[25] However, the above acoustic geometric-phase modulation usually requires a cumbersome helical waveguide,[26] which is not easy to integrate and thereby hinders their real applications.

Different from the acoustic metasurfaces that rely on exquisite resonance or the propagating phase shift achieved with complicated impedance[27-32] or effective refractive index engineering,[33-37] in this work, we propose a new kind of acoustic meta-device, namely, an acoustic geometric-phase meta-array based on a near-field coupled OAM transfer process, where a continuous acoustic geometric-phase modulation is available by simply varying the orientation angle of the acoustic element, e.g., a polar phase gradient metasurface (PGM) determining the TC conversion. Notably, our geometric-phase meta-atom can be designed to the deep subwavelength footprint, which endows our meta-arrays with better acoustic field reconstruction performance, showing great advantage in constructing high-quality acoustic meta-holograms. Then, acoustic geometric-phase meta-holograms made of pixelated array with adjusted amplitude and reconfigurable geometric-phase modulation are demonstrated, and TC-multiplexed acoustic holographic field reconstruction functionality is also investigated, which extends the information capacity of acoustic meta-holograms and shows good potential in constructing multifunctional and reconfigurable acoustic meta-devices.

## II. MATERIALS AND METHOD

Acoustic geometric-phase meta-array opens up a new avenue for reconfigurable acoustic wave control, where OAM transfer process plays a vital role in obtaining continuous phase modulation of the acoustic waves. Generally, cylindrical waveguide that supports the transport of acoustic vortices with specified TC is a good solution to acquiring the robust OAM transfer process. However, the nature of the existence of cut-off frequency inherently determines the minimal lateral size of the waveguide for target operating wavelength band, and thereby the pixel feature size is usually comparable to or even larger than the operating wavelength when high-order acoustic geometric phases are applied. Moreover, the implementation of cascaded vortex metasurfaces with wavelength-comparable intervals in geometric-phase meta-atom design further makes the whole device bulky. Here, we circumvent these problems by utilizing the near-field coupling between an evanescent acoustic vortex source and only one vortex metasurface, in which case the lateral size of the geometric-phase meta-atom could be compressed to a deep subwavelength scale, and the short coupling distance between the evanescent vortex source and metasurface greatly improves the compactness of acoustic geometric-phase meta-devices.

**A. Mechanism of acoustic geometric-phase via near-field coupled OAM transfer**

Since the information carried by an acoustic vortex is substantially characterized by its unique spiral phase front, the OAM transfer process could be straightforward understood as the switch between different spiral phase fronts. In this work, the key to obtaining acoustic geometric meta-atom of deep subwavelength footprints is to generate a perfect evanescent acoustic vortex that possesses a nice spiral phase front. Then we encode the acoustic geometric phase through the OAM transfer process between such an evanescent vortex of TC $l^{in}$ and an output plane wave of TC 0.

As a proof of concept of our near-field OAM transfer induced acoustic geometric phase, first, an acoustic phased transducer array (PTA) is placed in a short rigid cylindrical waveguide to generate a perfect evanescent acoustic vortex source of TC $l^{in}$, and the supporting panel around the point sources could be selected as the sound rigid wall (in this work) or sound absorbing materials. Next, a PGM of azimuthal period $l$ and orientation angle $\varphi_l$ couples out the evanescent acoustic field, where the azimuthal period is defined according to the angle period $a_\varphi$ in the polar direction, i.e., $a_\varphi = 2\pi/l$, and the absolute value of TC of the illuminating evanescent vortex field is also equal to $l$. Fig.1 (a) illustrates the schematic of acoustic field generations based on the PTA-PGM geometric-phase meta-atoms, where the spatial variation of the rotational angle of PGM could precisely modulate the phase of each pixel and the shape of the secondary wavefront.

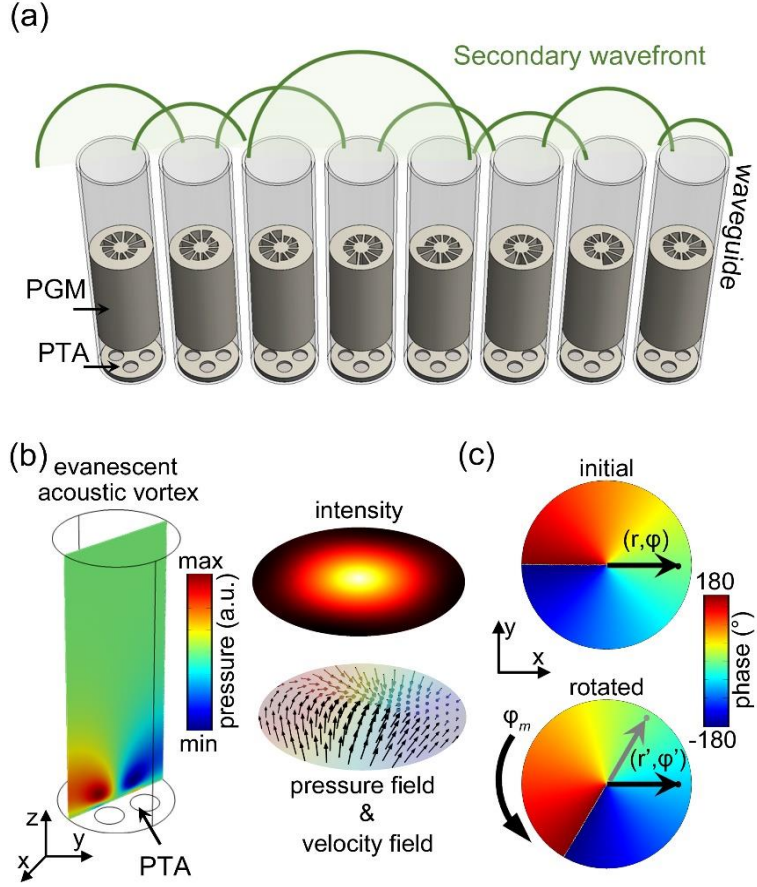

Fig. 1. (a) Schematic of acoustic field manipulation based on acoustic geometric-phase meta-atoms relying on near-field coupled OAM transfer, where spatially rotated PGM couples out a planar acoustic wave carrying the geometric phase. (b) Typical evanescent vortex field generated by PTA. (c) The phase distribution of an evanescent acoustic vortex of TC $l = 1$ before (top) and after (bottom) rotation by an angle of $\varphi_m$.

The general theory of the acoustic geometric phase obtained via near-field coupled OAM transfer process between an evanescent acoustic vortex and a vortex metasurface is given below.

It is well known that the pressure vortex field $p(r, \varphi, z)$ in a rigid cylindrical waveguide with radius $R$ can be expressed as a sum of eigenmodes:

$$p(r, \varphi, z) = \sum_m \sum_n P_{m,n} J_m(k_{m,n} r) \exp(jm\varphi) \exp(jk_z z), \qquad (1)$$

where r, φ and z are spatial coordinate in cylindrical waveguide; $P_{m,n}$ is the amplitude of $(m, n)th$ eigenmode, i.e., $J_m(k_{m,n}r) exp(jm\varphi) exp(jk_z z)$; $J_m(k_{m,n}r)$ is the $m$-th order cylindrical Bessel function, $m$ ($m = 0, \pm1, \pm2 \cdots$) is the TC of the eigenmode, $k_{m,n} = \alpha_{m,n}/k_0 R$ and $k_z = \sqrt{k_0^2 - k_r^2}$ are radial and axial wavenumber, and $\alpha_{m,n}$ is the $n$-th ($n = 0, 1, 2 \cdots$) solution of the expression $dJ_m(k_{m,n}r)/d(k_{m,n}r)\big|_{r=R} = 0$, and we generally consider $n = 0$ in our geometric-phase meta-devices design. When $\alpha_{m,n} > k_0 R$, the propagating constant $\beta_{m,n} = k_z/k_0$ is a pure imaginary number, therefore, the corresponding pressure field is an evanescent acoustic vortex but still holds an ideal spiral phase front (see Fig. 1(b)):

$$p_m(r, \varphi, z) = J_m(k_{m,n}r)\exp(jm\varphi)\exp(-|k_z|z). \tag{2}$$

According to Eq. (2), we see that the evanescent vortex maintains its spiral phase property (no propagation phase exists here) while its amplitude undergoes the exponential decay along $+z$ direction, therefore, it is possible to tune the amplitude of the coupled-out acoustic wave of the PTA-PGM system by setting proper coupling distance.

For the convenience of analysis, we denote the evanescent acoustic vortex given in Eq. (2) by $|m\rangle\exp(-|k_z|z)$, and the transmission operator of PGM is denoted by $\hat{T}(\theta_i)$, where $\theta_i$ refers to the initial orientation angle of PGM. Then the coupled-out plane acoustic wave $|0\rangle$ obtained via the near-field coupled OAM transfer process is given as:

$$|0\rangle = \hat{T}(\theta_i)|m\rangle\exp(-|k_z|d), \tag{3}$$

here $d$ is the coupling distance between the PTA and the PGM. When we rotate the PTA-PGM system counterclockwise by an angle of $\varphi_m$, then, the rotated coordinate (with apostrophe) is given by $r' = r$, $\varphi' = \varphi - \varphi_m$ and $d' = d$ (see Fig. 1 (c)), therefore, Eq. (3) turns to:

$$|0\rangle = \exp(-jm\varphi_m)\hat{T}(\theta_i + \varphi_m)|m\rangle\exp(-|k_z|d). \tag{4}$$

By combining Eq. (3) and (4), we have:

$$\hat{T}(\theta_i + \varphi_m) = \exp(jm\varphi_m)\hat{T}(\theta_i). \tag{5}$$

Eq. (5) indicates that the transmission operator of PGM would encode additional phase term of $\exp(jm\varphi_m)$ by rotating the structure. In addition, we can also obtain the above geometric phase term based on a general equation, i.e., $\exp[i(l^{in} - l^{out})\varphi_l]$ with

$l^{out}$ fixed as 0.[38]

**B. Properties of the acoustic geometric phase via near-field coupled OAM transfer**

In this work, we use the commercial finite element solver COMSOL Multiphysics to study the acoustic geometric-phase modulation retrieved from the near-field coupled OAM transfer processes. Here, a PTA made of several phased transducers is utilized to generate a high-quality evanescent vortex, which is robust against the interaction with the adjacent rotatable acoustic structure. Fig 2(a) shows four typical phase field distribution of the cross section of an evanescent vortex generated by PTA, where a good spiral phase front of designated TC $l = 1 \sim 4$ is correspondingly illustrated, here the operating wavelength is selected as 12 cm, the suggested phases of the transducers are labeled besides the colored solid circles.

A tunable spatial phase shift is available by adjusting the orientation angle of either PTA or PGM that composes the geometric-phase meta-atom. This working principle shows great advantage in reconfigurable acoustic field reconstruction when compared with devices that merely rely on the resonance or propagating phase (generally rely on dispersion design via tuning the geometric shape and size). In addition, the evanescent nature of the illuminating vortex endows our geometric-phase meta-atom with the feature of a deep subwavelength footprint, i.e., the operating frequency could be far below the cut-off frequency while effective near-field coupling can still be achieved. And the amplitude of the coupled-out plane wave can be compensated by adjusting the coupling distance $d$ between the PTA and PGM or tuning the drive voltage of the PTA.

For the simplicity of analysis and saving computing resources, we first utilize an ideal PGM for the demonstration of an arbitrary order acoustic geometric phase, see Fig.2 (b). The number of sectors in one polar period is $Q$, and each sector is a rigid waveguide filled with homogeneous and impedance-matched medium with the gradient acoustic refractive index, which aims for the graded phase delay in the azimuthal direction and maintains high transmittance. Assuming the height of PGM is $h$, and the operating wavelength is $\lambda_0$, then the refractive index step $\Delta n$ among neighboring sectors in one period is given as $\lambda_0/Qh$.

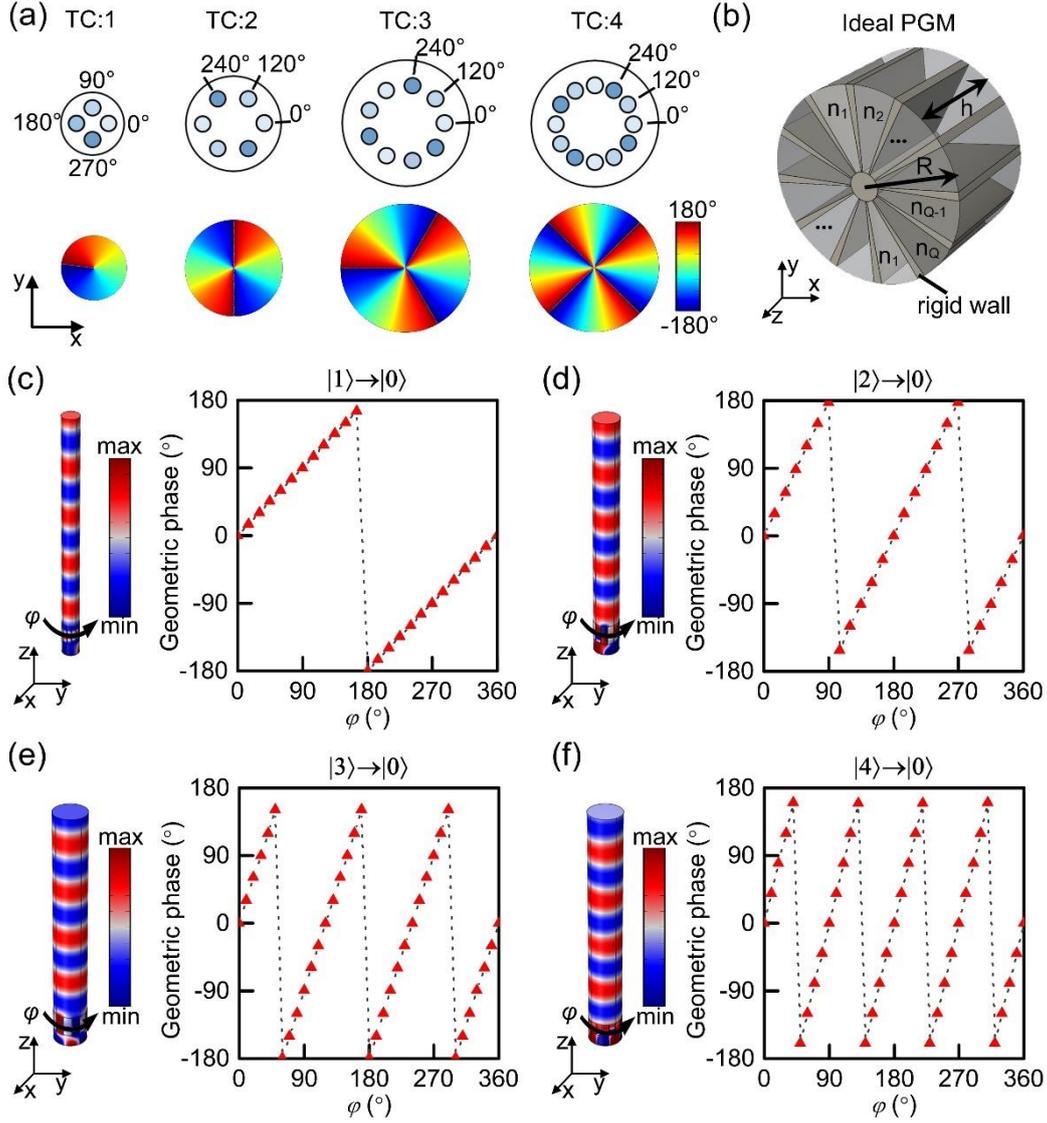

Fig. 2. TC-dependent acoustic geometric phase obtained via near-field coupled OAM transfer. (a)The phase field of the evanescent vortices generated by different PTAs. The radius of the transducer is 5 mm, the phase plane is sampled at a distance of 3 cm above the PTA, and the initial phase shifts of each transducer are labeled beside them. (b)Schematic of an ideal PGM. The radius R is the same as that of the waveguide, and the height $h$ of the PGM is selected as $0.5\lambda_0$. Geometric phases obtained via (c)$|1\rangle \to |0\rangle$, (d)$|2\rangle \to |0\rangle$, (e)$|3\rangle \to |0\rangle$ and (f)$|4\rangle \to |0\rangle$ near-field coupled OAM transfer processes, which are proportional to $\varphi$, $2\varphi$, $3\varphi$ and $4\varphi$, respectively.

Fig. 2 (c~f) show four typical cases of the acoustic geometric phase obtained by coupling the evanescent vortex of TC $l^{in} = l$ ($l = \pm 1, \pm 2 \cdots$) with an ideal PGM of TC $l^\xi$ which satisfies $|l^\xi| = |l|$, and the expected geometric-phase modulation is $\exp(il\varphi_l)$. Here, the transducer (color circle in Fig. 2 (a)) is regarded as the piston source with a designated initial phase and it is modeled as a vibrating surface with assigned normal acceleration in our simulations. In addition, the amplitude of normal acceleration can be tuned through the drive voltage of the transducer in real implementation, which could further modify the field intensity of the evanescent vortex and the coupled-out plane wave. The operating wavelength $\lambda_0$ is selected as 12 cm, and the coupling distance $d$ between the PTA and the PGM is fixed at 3 cm. The diameters of the waveguides utilized in Fig. 2(c), 2(d) and 2(e, f) are 4 cm, 6 cm and 8 cm, respectively, which are typical subwavelength values and much smaller than the minimal diameters that are required to support the first four vortex eigenmodes.

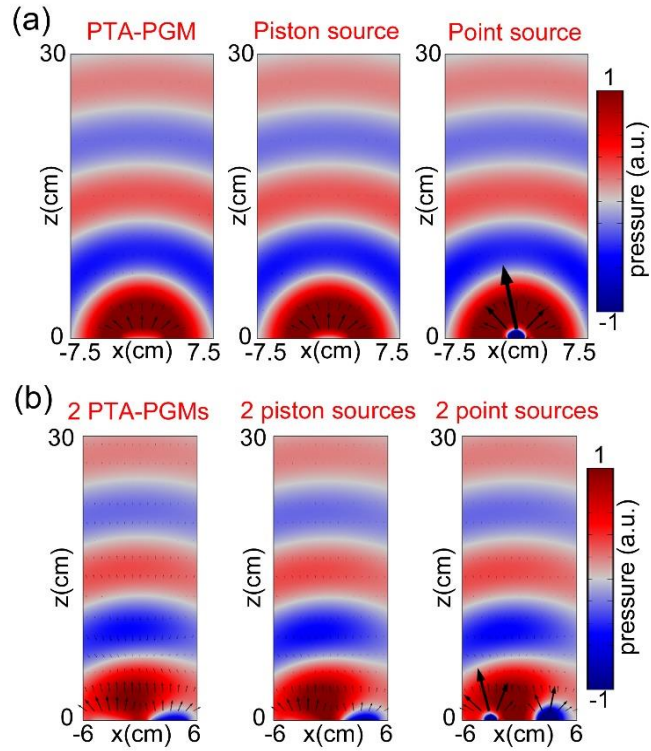

Fig. 3. Acoustic pressure and particle velocity field generated by (a) a PTA-PGM meta-atom based on the $|1\rangle \rightarrow |0\rangle$ near-field OAM transfer process, a piston source and a monopole point source, and (b) by two PTA-PGM meta-atoms, two piston sources and two monopole point sources. The rotation angles of the two PTA-PGM meta-atoms are 0° and 30°, respectively.

Interestingly, in this work, the PTA-PGM meta-atom behaves exactly the same as either an ideal piston source or monopole point source. Fig. 3(a) shows the pressure and velocity field of an individual real PTA-PGM supporting the $|1\rangle \rightarrow |0\rangle$ near-field OAM transfer process (left), piston source (middle) and monopole point source (right). We can observe nearly the same far-field pattern achieved with these three setups, and fields radiated by the PTA-PGM meta-atom and piston source even have almost the same velocity field distribution. Next, we consider the interaction between neighboring units. Fig. 3(b) shows the field generated by two close PTA-PGM meta-atoms (left) whose rotational angles are 0° and 30°, two piston sources (middle) and two monopole point sources (right) of designated initial phases of 0° and 30°. Here, the distance between two sources is $0.5\lambda_0$, which is a typical subwavelength value. Apparently, the far-field patterns of these two situations are almost the same. Therefore, we could simplify the full-wave simulation of the complex acoustic field generated with a large area geometric phase array by equivalently treating them as either piston sources (e.g., meta-lenses) or monopole sources (e.g., meta-holograms) for the purpose of saving computing resources.

**C. TC-multiplexed acoustic beam bending and focusing**

Interestingly, the acoustic geometric phase obtained via near-field OAM transfer could flip its sign by inverting the sign of the TC of the illuminating vortices while the transmitted amplitude remains almost unchanged. This intriguing property is similar to that of the optical geometric phase, which offers possibilities to enrich acoustic wave manipulation by utilizing the TC as a new degree of freedom. Fig. 4(a) shows TC-dependent beam bending realized with ideal PTA-PGM meta-atoms, where the deflected angle of the acoustic beam would flip its sign when inverting the incident vortex from $l^{in}$ to $-l^{in}$. As a proof of principle, we utilize the $|\pm 2\rangle \to |0\rangle$ near-field OAM transfer process to build a gradient geometric phased array. Here, the operating wavelength is 12 cm, the period is 6.5 cm, and 9 meta-atoms are gradually rotated by an angle of 20° to form the surface phase gradient. The anomalous refracted angles retrieved from the full-wave simulation are $-11.9° \pm 0.1°$ and $11.9° \pm 0.1°$, respectively, which agrees well with the theoretical values of $-11.84°$ and $11.84°$ when $l^{in}$ is switched from 2 to $-2$.

Similarly, considering the conjugate of the phase front would also turn a focusing lens to defocusing lens. The converged phase front of the form $\theta(x) = -k_0(f_z - \sqrt{x^2 + f_z^2})$ obtained with $|2\rangle \to |0\rangle$ near-field OAM transfer processes just switches to a diverged form of $\theta^*(x) = k_0(f_z - \sqrt{x^2 + f_z^2})$ obtained with $|-2\rangle \to |0\rangle$ process, where $k_0 = 2\pi/\lambda_0$ is the operating wavenumber, and the focal length $f_z$ is set as $12\lambda_0$. Twenty-nine meta-atoms are utilized in this acoustic geometric-phase metalens, and the rotation angle of each geometric-phase meta-atom is defined by $\varphi_l(x) = \theta(x)/|l^{in}|$. Fig. 4(b) shows the sound pressure level (SPL) of the acoustic field when the focusing and defocusing geometric metalenses are actively switched by flipping the sign of input evanescent vortex sources.

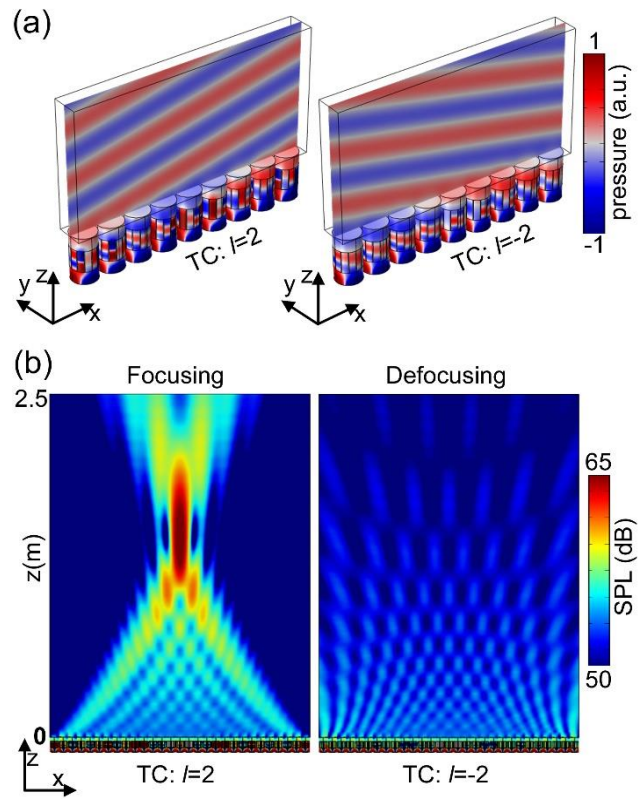

Fig. 4. (a)TC-dependent beam refraction and (b)focusing to defocusing switch based on the acoustic geometric meta-lens realized by $|\pm2\rangle \to |0\rangle$ near-field OAM transfer processes.

In analogy to the spin-multiplexed functionality demonstrated by optical GMPs, we further consider TC as a new degree of freedom in increasing the information processing capacity of acoustic meta-devices. In this section, we demonstrate one strategy of obtaining a multiple beam focusing functionality by using the supercell design, where two sets of geometric-phase meta-atoms independently account for two focusing functionalities, and up to three types of beam-focusing functionalities can be realized.

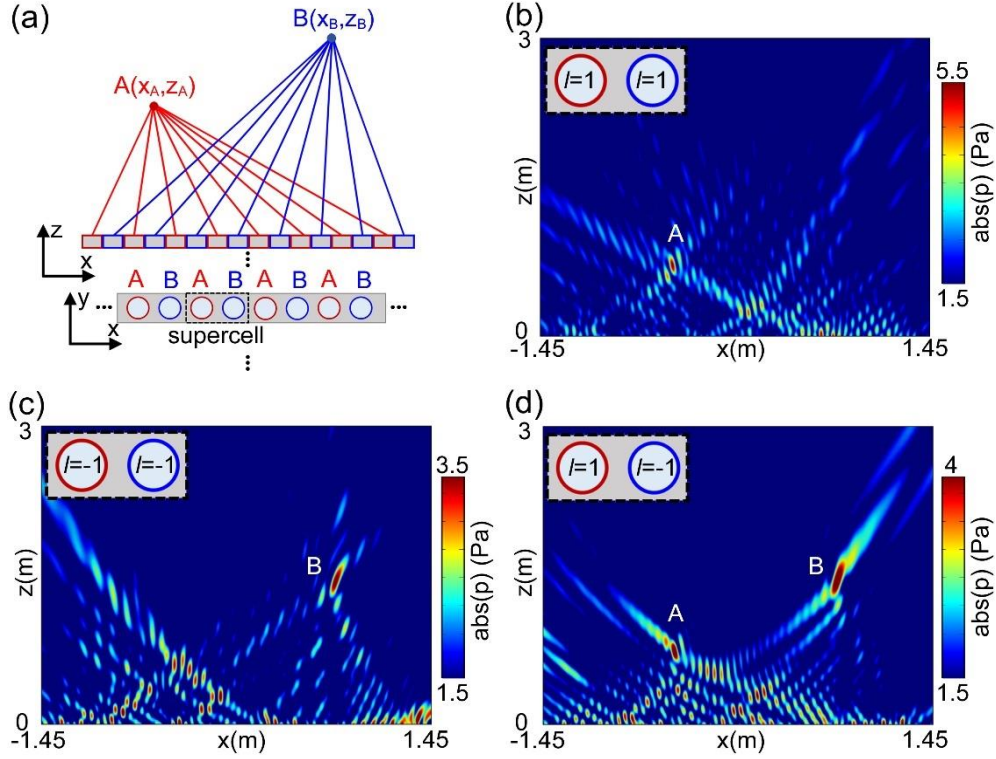

Fig. 5. TC-multiplexed beam focusing based on geometric-phase meta-atoms. (a)Schematic of TC-multiplexed beam focusing. Beam focusing realized by the (b)$|1\rangle \to |0\rangle$, (c) $|-1\rangle \to |0\rangle$, and (d)$|\pm1\rangle \to |0\rangle$ near-field coupled OAM transfer process. Red and blue circle refer to set A and B, respectively.

Fig.5(a) schematically demonstrates TC-multiplexed focusing by utilizing a $|\pm1\rangle \to |0\rangle$ near-field coupled OAM transfer process. The blue and green trajectories indicate the focusing of acoustic energy at points A and B via the $|1\rangle \to |0\rangle$ and $|-1\rangle \to |0\rangle$ near-field coupled OAM transfer processes, respectively. Here, the target operating wavelength is 12 cm, the size of the supercell is 5 cm×10 cm, and the distance between two subunits is 5 cm. The expressions of phase front of two converging meta-lenses are $\theta_A(x) = -k_0\left(f_A - \sqrt{(x-\Delta X_A)^2 + f_A^2}\right)$ and $\theta_B(x) = -k_0\left(f_B - \sqrt{(x-\Delta X_B)^2 + f_B^2}\right)$, which are correspondingly encoded with subunit sets A and B. In our design, $f_A = 6\lambda_0$, $f_B = 12\lambda_0$, $\Delta X_A = -5\lambda_0$, $\Delta X_B = 8\lambda_0$, and 40 supercells are used. By switching the TC $l^{in}$ of the illuminating evanescent vortex from 1 to $-1$, the position of focus switches from spatial position A $(-5\lambda_0, 6\lambda_0)$ to another position B $(8\lambda_0, 12\lambda_0)$. In addition, simultaneous observation of two focusing spots can be realized by separately controlling the TC of PTA of two sets of meta-atoms, i.e., set A and B, as shown in Fig. 5(a). However, it is also possible to obtain TC-multiplexed beam generation functionality without utilizing a supercell, see the following discussion on TC-multiplexed acoustic geometric-phase meta-holograms.

### D. Verification of the acoustic geometric phase with realistic PGM

As a proof of concept of our PTA-PGM geometric-phase meta-atom, we next illustrate the acoustic geometric-phase modulation achieved with realistic PGM made of pipe-Helmholtz resonator hybrid structures.[19] Fig. 6(a) shows the cross section of a fundamental unit (one layer) that accounts for a specific local phase shift. Here, the operating wavelength is selected as 12 cm, and 12 subunits are utilized to construct realistic PGMs that contain a specific azimuthal period number $l$. The height $h$ of the PGM is 7.5 cm, the radius of the PGM is $R$, the width of the straight pipe is $w_o$, and the widths of the open neck and Helmholtz resonator cavity are $w_o$ and $h_1$, respectively. The thickness $th_1$ of the rigid wall is 1.5 mm, and a solid central column of radius $R_c = 5$ mm is utilized to connect the subunits.

By evaluating the transmission coefficients of the subunit with varied open neck width and straight pipe width, we select the subunits of proper geometry parameters which possess the desired phase shift as well as high transmission efficiency (see Fig. 6(b)). In this section, two PGMs, whose radii are 2 cm and 3.5 cm, are designed for the $|1\rangle \rightarrow |0\rangle$ (one layer unit) and $|2\rangle \rightarrow |0\rangle$ (two-layer unit) near-field coupled OAM transfer processes, respectively (see the left panel of Fig. 6(c, d)). Considering the real application scenarios, we further mount the realistic PTA-PGM meta-atom on a rigid wall, its transmission property is retrieved from an infinite periodic array of geometric-phase meta-atoms (see the middle panel of Fig. 6(c, d)), and the periods are 2.5 cm and 4 cm, respectively. The distance between the PTA and PGM is 3 cm, and the distance between the PGM and rigid panel can be arbitrary. Here, we assume that the PGM directly couples out the evanescent vortex to free space without further propagating in the waveguide. Fig. 6(c, d) show the acoustic geometric phase obtained with realistic PGMs that correspondingly support the $|1\rangle \rightarrow |0\rangle$ and $|2\rangle \rightarrow |0\rangle$ near-field coupled OAM transfer processes, whose geometric-phase modulations are proportional to $\varphi$ and $2\varphi$, respectively.

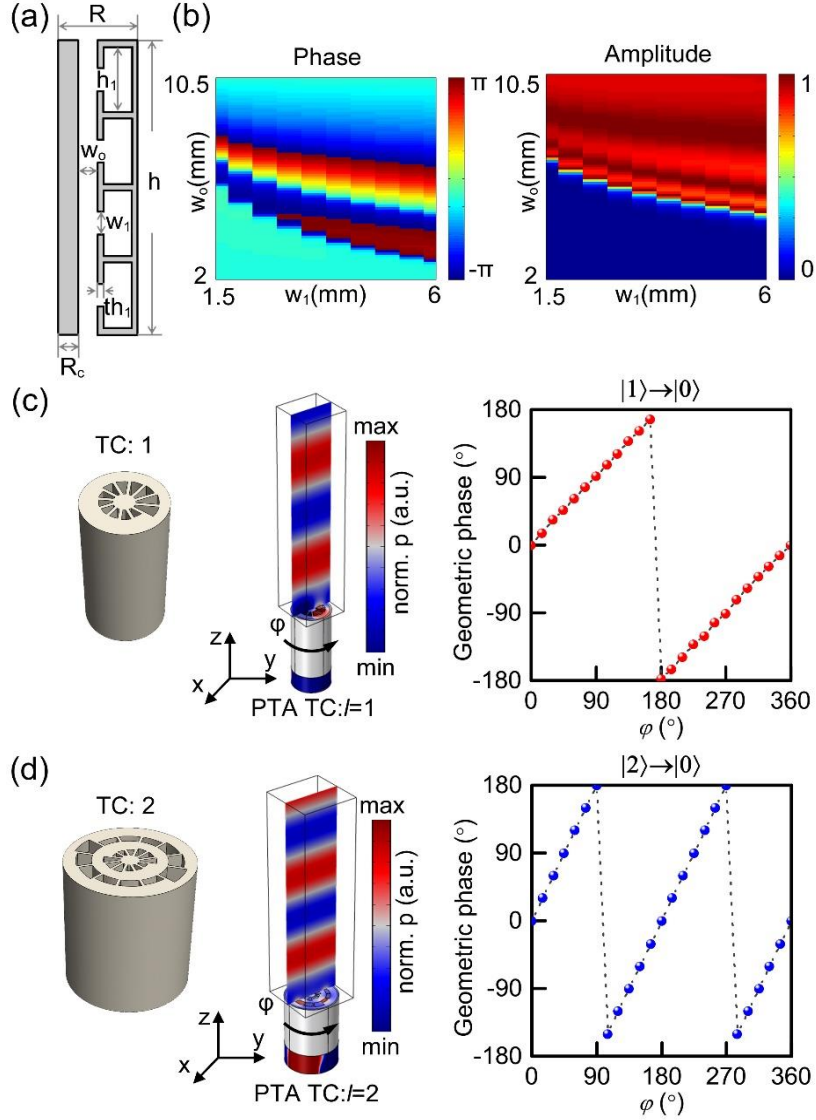

Fig. 6. Acoustic geometric phase obtained with realistic structures. (a) Cross section of a typical cylindrical hybrid pipe-Helmholtz structure. (b) Phase and amplitude of the transmission coefficient obtained by varying the width of the open neck of cavity $w_1$ and width of straight pipe $w_o$. The acoustic geometric phase modulation obtained through (c)$|1\rangle \rightarrow |0\rangle$ and (d)$|2\rangle \rightarrow |0\rangle$ near-field coupled OAM transfer processes, which are proportional to $\varphi$ and $2\varphi$.

## III. RESULTS

Acoustic meta-holograms have attracted great interest for the privilege they provide in obtaining complex acoustic fields in the target zone. Recently, acoustic meta-holograms have been realized with either resonance[39-43] or propagating phases.[44-48] However, previous studies on acoustic meta-holograms face the dilemma of non-reconfigurability and low information capacity. Here, acoustic geometric-phase meta-hologram (AGPMH) composed of PTA-PGM geometric-phase meta-atoms shows advantages in reconfigurable and TC-multiplexed multiple acoustic field reconstruction.

**A. Acoustic geometric-phase meta-hologram**

In this section, we exploit the potential of AGPMH for projecting a target acoustic holographic image at a desired distance. The phase and amplitude distribution of the AGPMH are obtained via the time-reversal method.[40,49] Generally, the amplitude modification of PTA-PGM geometric-phase meta-atoms is available by means of either varying the drive voltage of each PTA or modulating the coupling distance between the PTA and PGM. Considering the simplicity of implementation and future multifunctionality design, we assume that amplitude modification is achieved by tuning the amplitude of the drive voltage of PTAs.

The key to obtaining the acoustic meta-hologram is to first solve the pressure field distribution near the output surface of the meta-hologram from the target acoustic image. Then the meta-atoms supporting the same transmissive amplitude and phase modulations are utilized as the pixels. Here, the target image is divided into M×N pixels, and each pixel serves as a point source. For the ($m$, $n$)-th pixel, the acoustic field radiated by this point source is defined by its amplitude $A_{mn}$ and initial phase $\varphi_{mn}$, i.e., $P(X_m, Y_n, z) = A_{mn}\exp(i\varphi_{mn})$. Then, the pressure field at the output surface of the meta-hologram is contributed by all pixels of the target image, and thereby the pressure field of the ($k$, $l$)-th pixel of the meta-hologram is expressed as:

$$p(x_k, y_l, 0) = \sum_{m,n}^{M,N} \frac{A_{mn}}{r_{mn}} \exp[i(k_0 r_{mn} + \varphi_{mn})], \tag{6}$$

where $r_{mn} = \sqrt{(x_k - X_m)^2 + (y_l - Y_n)^2 + Z_d^2}$ is the distance between the image pixel ($m$, $n$) and hologram pixel ($k$, $l$), and $Z_d$ is the distance between AGPMH and the target image plane (we assume the AGPMH is at $z = 0$). Based on the time-reversal symmetry, the holographic image is given as:

$$P(X, Y, Z_d) = \sum_{k,l}^{K,L} \frac{A_{kl}}{r_{kl}} \exp[-i(k_0 r_{kl} - \varphi_{kl})], \tag{7}$$

where $r_{kl} = \sqrt{(X - x_k)^2 + (Y - y_l)^2 + Z_d^2}$, $A_{kl}$ and $\varphi_{kl} = l^{in}\theta_{kl}$ are the amplitude and geometric phase of the acoustic field generated by the meta-atom at the ($k$, $l$)-th pixel. Moreover, $l^{in}$ is the TC of the evanescent acoustic vortex generated by PTA, $\theta_{kl}$ is the rotational angle of the PGM of the meta-atom at the ($k$, $l$)-th pixel.

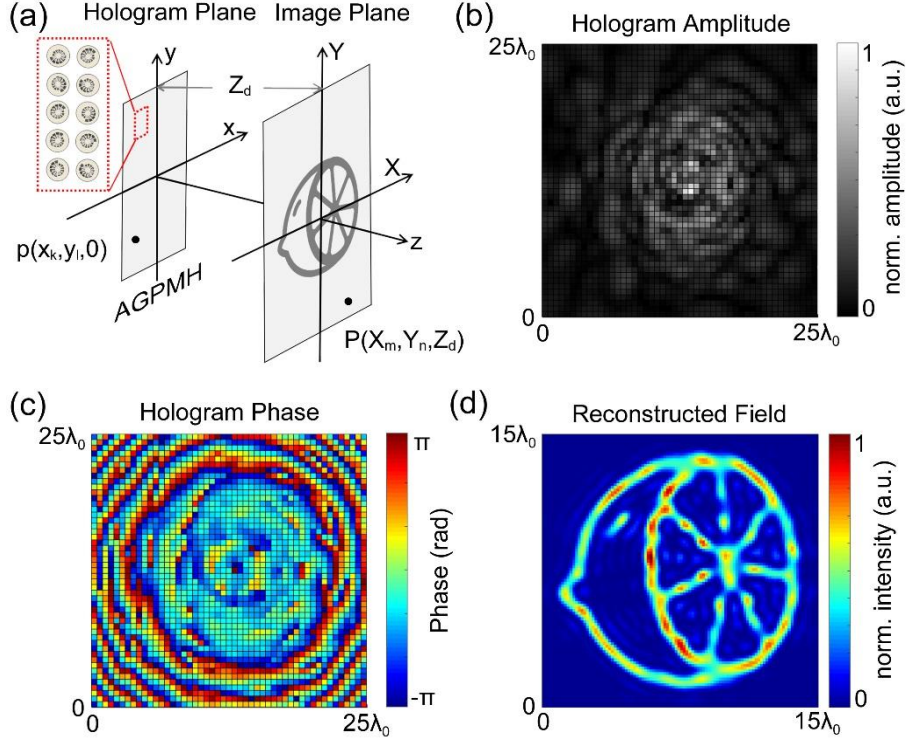

Fig. 7. Acoustic meta-hologram realized with acoustic geometric-phase meta-atoms. (a) Schematic of working principle of AGPMH. (b) Designed amplitude and phase of a complex AGPMH. (c) Numerical reconstruction of the target acoustic field based on the designed AGPMH.

Fig. 7(a) shows the working principle of our AGPMH. For each pixel, the amplitude of coupled-out acoustic plane waves is normalized with respect to the maximum amplitude of the designated AGPMH, which can be precisely tuned via the drive voltage of the PTA. Phase modulation is achieved with the rotation of the PGM (see the red dashed box shown in Fig. 7(a)), which can be continuously tuned from 0° to 360°, and continuous phase allowed by acoustic geometric phase can facilitate better holographic field quality. Fig. 7(b, c) show the amplitude and phase distribution of an AGPMH composed of 50×50 units, here the pixel size is $0.5\lambda_0$ and the target image is reconstructed at a distance of $10\lambda_0$. Based on our previous discussion given in Fig. 3, the geometric PTA-PGM meta-atom is equivalently treated as a point source, and Fig. 7(d) shows the reconstructed target acoustic field based on a cluster of point sources modulated with the amplitude and phase shown in Fig. 7(b, c), which agrees well with the expected "Lemon" pattern.

## B. Multiplane acoustic geometric-phase meta-hologram

Next, we investigate the multiplane acoustic meta-hologram achieved with our subwavelength geometric-phase meta-atoms. Considering the design procedure given in Section 3.1, we only need to add the collective contribution of the pixels of the j-th target image by rewriting Eq. 6 as:

$$p(x_k, y_l, 0) = \sum_j^J \sum_{m,n}^{M,N} \frac{A_{mn}}{r_{mn}} \exp[i(k_0 r_{mn} + \varphi_{mn})], \tag{8}$$

where J is the overall number of target image planes.

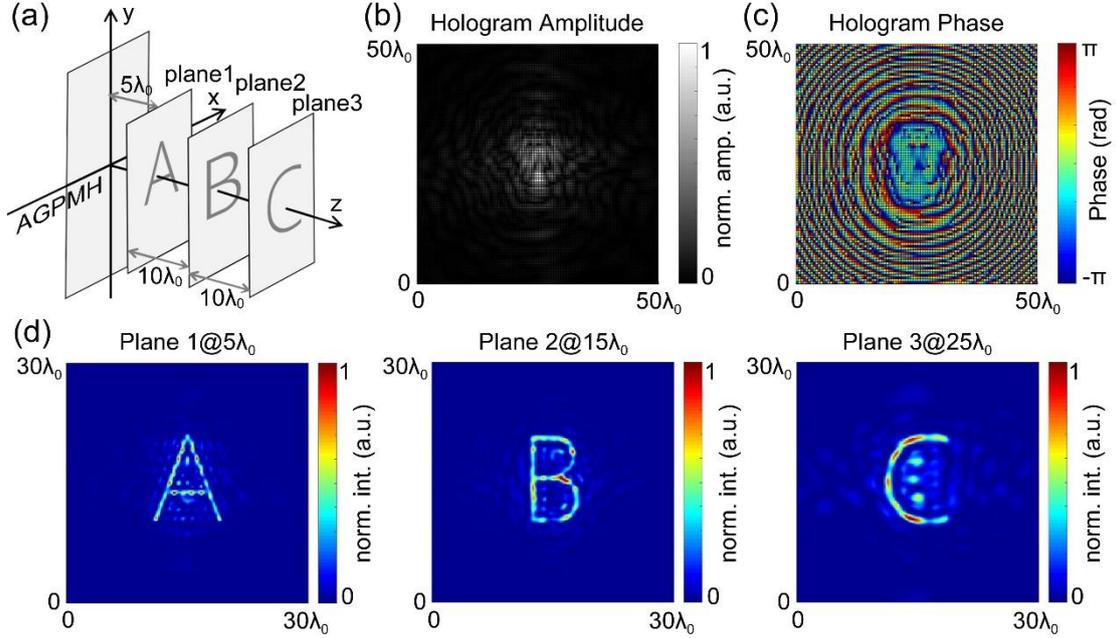

Fig. 8. AGPMH for designated acoustic field reconstruction at different depths. (a) Schematic of multiplane AGPMH. Designed (b)amplitude and (c)phase distribution of the complex meta-hologram. (d)Numerically reconstructed acoustic field of patterns "A", "B", and "C" at the distances of $5\lambda_0$, $15\lambda_0$ and $25\lambda_0$, respectively.

Fig. 8(a) gives the schematic description of our multiplane AGPMH, and three desired images of letters "A", "B" and "C" are reconstructed at different positions of $5\lambda_0$, $15\lambda_0$ and $25\lambda_0$ away from the meta-hologram sample. The designed complex meta-hologram is composed of 100×100 units, and the pixel size is $0.5\lambda_0$. Fig. 8(b, c) show the amplitude and phase of each composing pixel, and Fig. 8(d) gives the reconstructed acoustic field at the designed distance.

## C. TC-multiplexed acoustic geometric-phase meta-hologram

Apart from the above demonstrations, our PTA-PGM geometric-phase meta-atoms could also support TC-dependent geometric-phase modulation. Therefore, in this section, we address the potential of TC-multiplexed acoustic holographic field generation with AGPMH, which can greatly enhance its information capacity.

First, we analyze the influence of the inversion of the sign of the TC on the reconstructed acoustic fields. When we invert the TC of the evanescent acoustic vortex generated by the PTA from $l^{in}$ to $-l^{in}$, then the phase of the transmitted acoustic field near the AGPMH inverts its sign as well, and the holographic acoustic field given by Eq. (7) becomes:

$$P'(X, Y, z) = \text{conj}\left\{\sum_{k,l}^{K,L} \frac{A_{kl}}{r_{kl}} \exp[i(k_0 r_{kl} + l^{in}\theta_{kl})]\right\}, \tag{9}$$

where *conj* refers to the conjugate operator. The expression inside the conjugate operator exactly describes the acoustic field reconstructed by the hologram with amplitude distribution $A_{kl}(X_k, Y_l)$ and phase distribution $\varphi_{kl}(X_k, Y_l) = l^{in}\theta_{kl}(X_k, Y_l)$ towards the $-z$ direction, which is opposite to that utilized in Eq. (7). Therefore, when we invert the sign of TC, the meta-hologram is supposed to reconstruct a conjugate and virtual image at the opposite position, i.e., $z = -Z_d$.

Considering a hologram that could simultaneously reconstruct two independent holographic images located at opposite sides of the meta-hologram, e.g., a real image "Heart" and a virtual image "Star", see Fig. 9(a). The design of such meta-holograms can be realized by directly adding the field propagating from the target images located at $z = Z_{1,d}$ and $z = -Z_{2,d}$. Given two fields $P_1(X_m, Y_n, Z_{1,d}) = A_{1,mn}\exp(i\varphi_{1,mn})$ and $P_2(X_m, Y_n, -Z_{2,d}) = A_{2,mn}\exp(i\varphi_{2,mn})$, the overall field at hologram plane $z = 0$ is:

$$p(x_k, y_l, 0) = p_1(x_k, y_l, 0) + p_2(x_k, y_l, 0), \tag{10}$$

where

$$p_1(x_k, y_l, 0) = \sum_{m,n}^{M,N} \frac{A_{1,mn}}{r_{1,mn}} \exp[i(k_0 r_{1,mn} + \varphi_{1,mn})], \tag{11}$$

$$p_2(x_k, y_l, 0) = \sum_{m,n}^{M,N} \frac{A_{2,mn}}{r_{2,mn}} \exp[-i(k_0 r_{2,mn} - \varphi_{2,mn})]. \tag{12}$$

Here, $r_{j,mn} = \sqrt{(x_k - X_m)^2 + (y_l - Y_n)^2 + Z_{j,d}^2}$, $j = 1,2$.

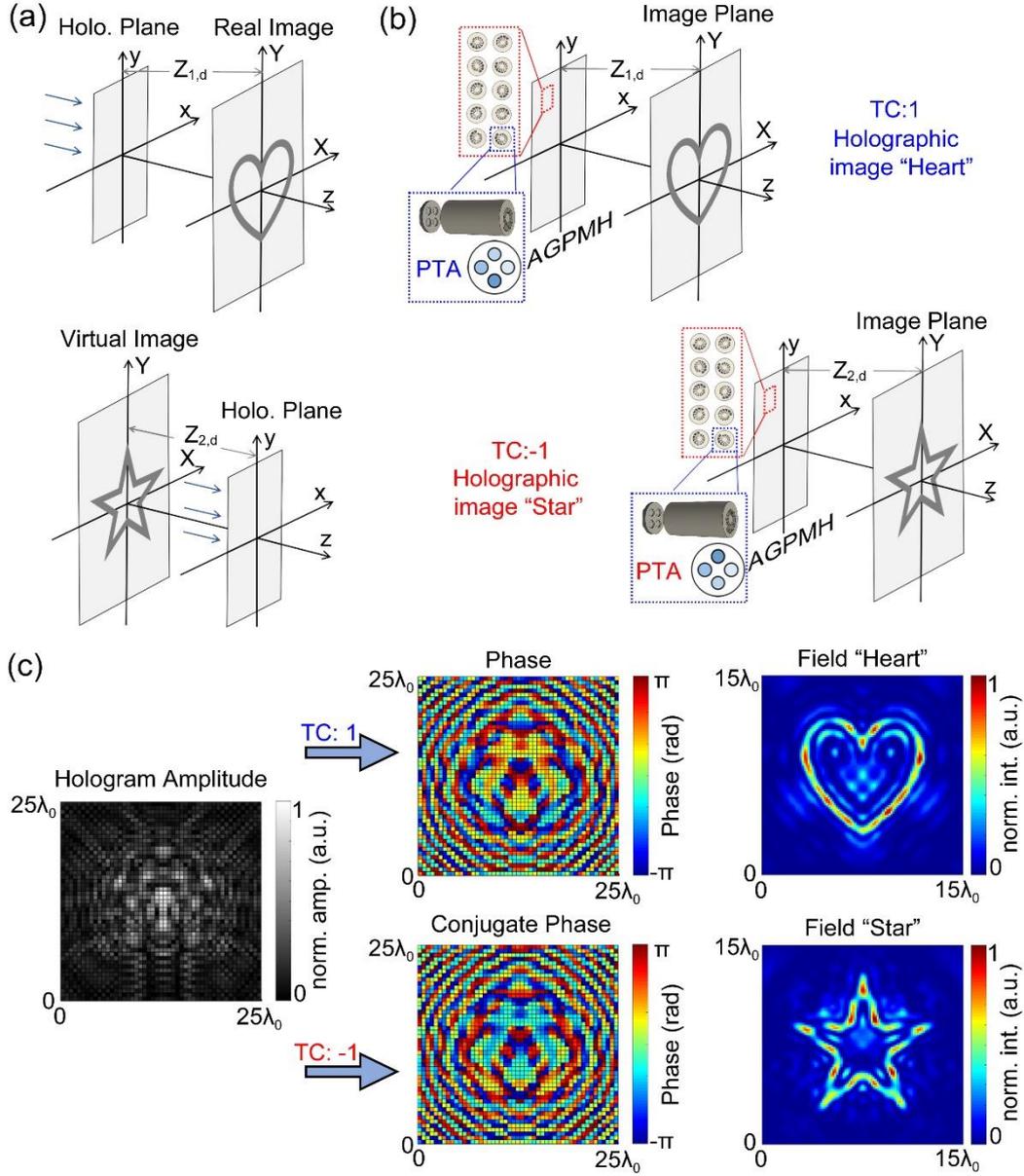

Fig. 9. TC-multiplexed AGPMH. (a) Classical hologram that could reconstruct two independent images at its front and back half spaces. (b) Schematic of TC-multiplexed AGPMH. When the TC of the evanescent vortex generated by PTA switches from 1 to -1, the reconstructed field switches from field "Heart" to field "Star". (c) Designed meta-hologram and reconstructed image based on TC-multiplexed geometric phases.

Interestingly, when we invert the sign of TC of illuminating evanescent vortex, the transmitted field near the surface of meta-hologram is:

$$p^*(x_k, y_l, 0) = p_1^*(x_k, y_l, 0) + p_2^*(x_k, y_l, 0), \qquad (13)$$

where $*$ refer to the conjugate. Based on the above analysis, then two identical target

**19 / 25**

images are supposed to appear, i.e., virtual image "Heart" at $z = -Z_{1,d}$ and real image "Star" at $z = Z_{2,d}$. Therefore, our AGPMH could selectively reconstructs different images at plane $z = Z_{1,d}$ or $z = Z_{2,d}$ as the TC of PTA varies its sign.

Fig. 9(b) shows the conceptual schematic of a TC-multiplexed AGPMH. Here, two acoustic holographic fields are available by flipping the sign of the evanescent vortex sources that are generated by PTAs, while the spatial orientation angles of all PGMs remain unchanged. Different from the TC-multiplexed functionality given in Section 2.3, here, every geometric-phase meta-atom simultaneously contributes to the two holographic pressure fields.

Fig. 9(c) shows the numerical reconstruction of the target acoustic image "Heart" and "Star" when flipping the sign of the evanescent vortex. Here, the phase distributions of AGPMHs designated for images "Heart" and "Star" are mutually conjugated, while the amplitude distributions are the same. The AGPMH shown in this section is composed of 50×50 units, the pixel size of the meta-hologram is $0.5\lambda_0$, and the target images are reconstructed at the same distance of $10\lambda_0$.

## IV. DISCUSSIONS AND CONCLUSION

The advantage of such strategy by utilizing AGPMH is that we could achieve continuous phased control by changing the rotational angle of either PTA or PGM while only few levels of phase delay of the drive electric signals are required (3 or 4 levels, see Fig. 2(a)), while classical transducer arrays designed for complex acoustic field reconstructions generally require cumbersome electrical delay lines in controlling both phase and amplitude of numerous signal channels.

In addition, we could construct a programmable AGPMH by rotating the PTA that is mounted on a step motor. Based on the proper selection of the order of acoustic geometric phase (the order of the geometric phase is related to the sensitivity of the phase variation to rotational angle and thereby the modulation speed), we could achieve rapid and high-resolution acoustic field reconstructions with this programmable AGPMH. It should be noted that if the evanescent acoustic vortex generated with PTA is available with a passive structure, our acoustic geometric-phase meta-atoms could be extended to work in the all-passive acoustic wave manipulation.

In summary, we utilize the near-field coupled OAM transfer process to realize effective acoustic geometric-phase modulation with a deep-subwavelength footprint. Based on the PTA, a nearly perfect evanescent acoustic vortex of the desired TC can be generated and further coupled with the PGM of the corresponding TC. Then, the plane acoustic wave transmitted from the PGM can carry tunable geometric phases by varying the orientation angle of either the PTA or PGM, and such acoustic geometric-phase modulation is successfully verified with realistic PGM structures. Thanks to the deep-subwavelength geometric size of our PTA-PGM meta-atoms and controllable sign of TC of evanescent acoustic vortex source, different TC-multiplexed devices, especially the AGPMH are comprehensively studied. Our work opens up a new possibility in high-quality acoustic field reconstructions and multifunctional acoustic meta-holograms.

## ACKNOWLEDGEMENTS

The kind help offered by Dr. Yuhong Na (Y) in polishing the language of the manuscript is acknowledged. This project was supported by the National Nature Science Foundation of China (Grant Nos. 12104044, 92050117, 61861136010), Beijing Outstanding Young Scientist Program (Grant No. BJJWZYJH0120190007022), China Postdoctoral Science Foundation (Grant No. 2021M690410) and Sino-German (CSC-DAAD) Postdoc Scholarship Program, 2020 (Grant No. 57531629).

## AUTHOR DECLARATIONS

**Conflict of Interest**

The authors have no conflicts to disclose.

## DATA AVAILABILITY

The data that support the findings of this study are available from the corresponding author upon reasonable request.
**22** / **25**

# References

1. N. Yu, P. Genevet, M. A. Kats, F. Aieta, J.-P. Tetienne, F. Capasso, and Z. Gaburro, Science **334**, 333-337 (2011).

2. S. Sun, Q. He, S. Xiao, Q. Xu, X. Li, and L. Zhou, Nat. Mater. **11**, 426-431 (2012).

3. L. Huang, X. Chen, H. Mühlenbernd, H. Zhang, S. Chen, B. Bai, Q. Tan, G. Jin, K.-W. Cheah, C.-W. Qiu, J. Li, T. Zentgraf, and S. Zhang, Nat. Commun. **4**, 2808 (2013).

4. G. Zheng, H. Mühlenbernd, M. Kenney, G. Li, T. Zentgraf, and S. Zhang, Nat. Nanotechnol. **10**, 308-312 (2015).

5. W. Zhao, H. Jiang, B. Liu, Y. Jiang, C. Tang, and J. Li, Sci. Rep. **6**, 30613 (2016).

6. J. P. B. Mueller, N. A. Rubin, R. C. Devlin, B. Groever, and F. Capasso, Phys. Rev. Lett. **118**, 113901 (2017).

7. D. Lin, P. Fan, E. Hasman, and M. Brongersma, Science **345**, 298-302 (2014).

8. M. Khorasaninejad, W. T. Chen, R. C. Devlin, J. Oh, A. Y. Zhu, and F. Capasso, Science **352**, 1190-1194 (2016).

9. S. Wang, P. C. Wu, V. -C. Su, Y. -C. Lai, M. -K. Chen, H. Y. Kuo, B. H. Chen, Y. H. Chen, T. -T. Huang, J. -H. Wang, R. -M. Lin, C. -H. Kuan, T. Li, Z. Wang, S. Zhu, and D. P. Tsai, Nat. Nanotechnol. **13**, 227-232 (2018).

10. E. Tseng, S. Colburn, J. Whitehead, L. Huang, S. -H, Baek, A. Majumdar, and F. Heide, Nat. Commun. **12**, 6493 (2021).

11. P. Genevet, F. Capasso, F. Aieta, M. Khorasaninejad, and R. Devlin, Optica **4**, 139-152 (2017).

12. C. Shi, R. Zhao, Y. Long, S. Yang, Y. Wang, H. Chen, J. Ren, and X. Zhang, 2019 Natl. Sci. Rev. **6** 707-712 (2019).

13. K. Y. Blikoh, and F. Nori, Phys. Rev. B **99**, 174310 (2019).

14. Y. Long, D. Zhang, C. Yang, J. Ge, H. Chen, and J. Ren, Nat. Commun. **11**, 4716 (2020).

15. I. Rondón, J. Phys. Commun. **5**, 085015 (2021).

16. S. Wang, G. Zhang, X. Wang, Q. Tong, J. Li, and G. Ma, Nat. Commun. **12**, 6125
**23** / **25**

(2021).

17. C. Shi, M. Dubois, Y. Wang, and X. Zhang, PNAS, **114**, 7250-7253 (2017).

18. X. Jiang, B. Liang, J. -C Cheng, and C. -W. Qiu, Adv. Mater. **30**, 1800257 (2018).

19. X. Jiang, Y. Li, B. Liang, J. -C. Cheng, and L. Zhang, Phys. Rev. Lett. **117**, 034301 (2016).

20. H. Esfahlani, and H. Lissek, Phys. Rev. B **95**, 024312 (2017).

21. X. Jiang, J. Zhao, S. -L. Liu, B. Liang, X. -Y. Zhou, J. Yang, C. -W. Qiu, and J. C. Cheng, Appl. Phys. Lett. **108**, 203501 (2016).

22. Z. Hou, H. Ding, N. Wang, X. Fang, and Y. Li, Phys. Rev. Appl. **16**, 014002 (2021).

23. Y. Fu, C. Shen, X. Zhu, J. Li, Y. Liu, S. A. Cummer, and Y. Xu, Sci. Adv. **6**, eaba9876 (2020).

24. S. Wang, G. Ma, and C. T. Chan, Sci. Adv. **4**, eaaq1475 (2018).

25. K. Y. Bliokh, Phys. Rev. Lett. **97**, 043910 (2006).

26. F. Liu, W. Li, Z. Pu, and M. Ke, Appl. Phys. Lett. **114**, 193501 (2019).

27. J. Zhao, B. Liang, Z. N. Chen, and C. -W. Qiu, Appl. Phys. Lett. **103**, 151604 (2013).

28. Y. Zhu, X. Fan, B. Liang, J. -C. Cheng, and Y. Jing, Phys. Rev. X **7**, 021034 (2017).

29. X. Wu, X. Xia, J. Tian, Z. Liu, and W. Wen, Appl. Phys. Lett. **108**, 163502 (2016).

30. A. Díaz-Rubio, J. Li, C. Shen, S. A. Cummer, and S. A. Tretyakov, Sci. Adv. **5**, eaau7288 (2019).

31. G. Ma, M. Yang, S. Xiao, Z. Yang, and P. Sheng, Nat. Mater. **13**, 873-878 (2014).

32. Y. Li, X. Jiang, R.-Q. Li, B. Liang, X. -Y. Zou, L. -L. Yin, and J. -C. Cheng, Phys. Rev. Appl. **2**, 064002 (2014).

33. S. Zhang, L. Yin, and N. Fang, Phys. Rev. Lett. **102**, 194301 (2009).

34. Z. Liang and J. Li, Phys. Rev. Lett. **108**, 114301 (2012).

35. Y. Li, B. Liang, X. Tao, X.-F. Zhu, X. -Y. Zou, and J. -C Cheng, Appl. Phys. Lett. **101**, 233508 (2012).

36. X. Zhu, K. Li, P. Zhang, J. Zhu, J. Zhang, C. Tian, and S. Liu, Nat. Commun. **7**, 11731 (2016).

37. Y. Tian, Q. Wei, Y. Cheng, Z. Xu, and X. Liu, Appl. Phys. Lett. **107**, 221906 (2015).

38. B. Liu, Z. Su, Y. Zeng, Y. Wang, L. Huang, and S. Zhang, New J. Phys. **23**, 113026 (2021).

39. Y. Xie, C. Shen, W. Wang, J. Li, D. Suo, B.-I. Popa, Y. Jing, and S. A. Cummer, Sci. Rep. **6**, 35437 (2016).

40. Y. Zhu, J. Hu, X. Fan, J. Yang, B. Liang, X. Zhu, and J. Cheng, Nat. Commun. **9**, 1632 (2018).

41. Y. Zhu, N. J. Gerard, X. Xia, G. C. Stevenson, L. Cao, S. Fan, C. M. Spadaccini, Y. Jing, and B. Assouar. Adv. Funct. Mater. **31**, 2101947 (2021).

42. S. Fan, Y. Zhu, L. Cao, Y. Wang, A. L. Chen, A. Merkel, Y. Wang, and B. Assouar, Smart Mater. Struct. **29**, 105038 (2020).

43. H. Zhang, W. Zhang, Y. Liao, X. Zhou, J. Li, G. Hu, and X. Zhang, Nat. Commun. **11**, 3956 (2020).

44. K. Melde, A. G. Mark, T. Qiu, and P. Fischer, Nature **537**, 518-522 (2016).

45. M. D. Brown, Appl. Phys. Lett. **115**, 053701 (2019).

46. J. Kim, S. Kasoji, P. G. Durham, and P. A. Dayton, Appl. Phys. Lett. **118**, 051902 (2021).

47. L. Cox, K. Melde, A. Croxford, P. Fischer, and B. W. Drinkwater, Phys. Rev. Appl. **12**, 064055 (2019).

48. S. Jiménez-Gambín, N. Jiménez, J. Benlloch, and F. Camarena, Phys. Rev. Appl. **12**, 014016 (2019).

49. M. Fink, D. Cassereau, A. Derode, C. Prada, P. Roux, M. Tanter, J. -L. Thomas, and F. Wu, Rep. Prog. Phys. **63**, 1933 (2000).